\documentclass{ws-procs9x6}

\newcommand{\be}{\begin{equation}}
\newcommand{\ee}{\end{equation}}
\newcommand{\ba}{\begin{eqnarray}}
\newcommand{\ea}{\end{eqnarray}}

\begin{document}

\title{LOOP QUANTUM GRAVITY INDUCED CORRECTIONS TO FERMION DYNAMICS IN FLAT SPACE}

\author{LUIS F. URRUTIA}
\address{Departamento de F\'\i sica de Altas Energ\'\i as \\
Instituto de Ciencias Nucleares\\
Universidad Nacional Aut\'onoma de M\'exico\\
Apartado Postal 70-543\\
04510 M\'exico, D.F. \\
E-mail: urrutia@nuclecu.unam.mx}

\maketitle

\abstracts{A summary of recent work related to the calculation of
loop quantum gravity induced corrections to standard particle
(photons and spin 1/2 fermions) dynamics in flat space is
presented.  Stringent bounds upon the parameters characterizing
the corrections in the fermionic sector, arising from already
known clock-comparison experiments, are reviewed. }

\section{INTRODUCTION}

Loop Quantum Gravity (LQG) has provided a highly successful
approach to deal with the old problem of constructing a quantum
version of Einstein general relativity. Among its predictions we
singularize the property that the area and volume operators are
quantized in the corresponding units of the Planck length $\ell_P$
.\cite{ROVSMOL} In this way, a continuum description of space
should be considered only as an approximation valid for length
scales $L >> \ell_P $. The granular structure of space at short
distances poses at least two interesting problems certainly tied
to each other. On one hand there is the question of whether or not
such granular structure will induce modifications to standard
particle dynamics in flat space, for example. A most promising
observational consequence of these corrections  is the
modification of the standard energy-momentum relations for
particles.\cite{AMC} The analogy with particle propagation in a
lattice crystal, for example, suggests that modifications will
indeed arise. Secondly, since that granular structure incorporates
the notion of a minimum (maximum) length (momentum), another
important related issue is how these corrections would affect
Einstein special relativity principle.

The resolution of the first question in LQG requires a detailed
and exact construction of semiclassical states that would
approximate the matter content plus the flat space geometry at
large distances, while retaining the granular structure at the
Planck scale. Progress in this direction has been
made,\cite{THIEMANN2} though the final answer still remains an
open problem. In this way, many of the estimations of the
corrections to particle dynamics arising from LQG are based upon
an heuristical approach that intends to capture what one would
expect to be the most relevant and general properties of such
semiclassical states.\cite{GP,URRUF,URRUAD}

With respect to the second question, we point out  some
alternatives that have been considered in the literature:

(1) The fact that such modifications lead to an energy dependent
velocity of light, for example, has normally been taken as an
indication that their mere existence would imply a violation of
the active (particle) form of the relativity principle expressed
in terms of the standard Lorentz transformations. This approach to
the problem makes direct contact with a large body of work
previously developed to describe the most general Lorentz
violating dynamics which is compatible with the standard model of
strong and electroweak interactions,\cite{kostelecky2} and more
recently with an Einstein-Cartan description of
gravity.\cite{kostelecky3} This standard model extension has
provided a unified framework to analyze most of the experimental
work designed to probe Lorentz violations in
nature.\cite{kostelecky4}

(2) A second approach has recently emerged which retains a full
relativity principle by extending or deforming  the standard
Lorentz transformations in a manner consistent with the induced
dynamical modifications, in particular with deformed dispersion
relations. These proposals are included under the generic name of
double special relativity (DSR), which typically incorporate
$\ell_P$ as a second invariant quantity in the same footing as the
speed of light in the low energy approximation.\cite{DSR}

(3) There is also the possibility that standard Lorentz covariance
is preserved.\cite{ROVSOR}

The paper is organized as follows: Section 2 contains a summary of
the heuristical approach  pursued to obtain the modified dynamics
arising from LQG for two-component spin $1/2$ fermions and
photons, respectively, in flat space.\cite{URRUF,URRUAD} Section 3
summarizes the bounds upon the parameters in the free fermionic
sector, obtained assuming the existence of a preferred reference
frame.\cite{boundsurru}

\section{MODIFIED DYNAMICS FOR FIELDS $F$ IN  FLAT SPACE }

Central to the approach of Alfaro, Morales-T\'ecotl and Urrutia
\cite{URRUF,URRUAD} is Thiemann's regularization of the LQG
Hamiltonians ${\hat H}_{\Gamma}$.\cite{ThiemannR} This is based
upon a triangulation of space, adapted to the corresponding graphs
$\Gamma$ which define a given state. The regularization is
provided by the volume operator, with discrete eigenvalues arising
only from the vertices of the graph. Here we take an heuristical
point of view, starting from the exact operator version of LQG and
defining its action upon the semiclassical state through some
plausible requirements. We think of the semiclassical
configuration describing a particular matter or gauge field
operator ${\hat F}$  plus gravity, as given by an ensemble of
graphs $\Gamma$, each occurring with probability $P(\Gamma)$. To
each of such graphs we associate a wave function $|\Gamma, L, F
\rangle$ which is peaked with respect to the classical  field
configuration $F$, together with a flat gravitational metric and a
zero value for the gravitational connection at large distances. In
other words, the contribution of the gravitational operators
inside the expectation value  is estimated as.\cite{URRUF,URRUAD}
\begin{eqnarray}
\langle\Gamma, {L}, F|\, ...{\hat q}_{ab}...\,|\Gamma, {L}, F
\rangle&=& \delta_{ab} + O\left(\frac{\ell_P}{ L}\right),
\nonumber \\
\langle\Gamma, { L}, F |\, ...{\hat A}_{ia}...\,|\Gamma, { L}, F
\rangle&=& 0\, + \frac{1}{ L}\, \left(\frac{\ell_P}{{
L}}\right)^\Upsilon. \label{EXPV}
\end{eqnarray}
The parameter $\Upsilon \geq 0$ is a real number.
 Also we associate the effective hamiltonian $H_{\Gamma}=\langle \Gamma,L,F\,|{\hat H}_{\Gamma}
 |\, \Gamma,L,F\rangle$  to each graph.
 The scale ${ L}>> \ell_P \, $ of the wave function is such that the continuous
 flat metric approximation is appropriate for distances much larger that $L$,
 while the granular structure of space becomes relevant when probing
 distances smaller that $L$. In this way, space is constructed by
 adding
 boxes of size $L^3$, which center represents a given point $\bf x$ in the
 continuum and which contain a large number of vertices of the adapted graphs.
 The  field $F$, characterized by a De Broglie wave length
 $\lambda$,
 is considered a slowly varying function within each box ( $\lambda >
 L$) and  contributes with  its classical value at the center
 of the box, when taking expectation values. Gravitational
 variables are rapidly varying inside the box.
The total effective Hamiltonian is defined as an  average over the
graphs $\Gamma$: ${\rm H}=\sum_{\Gamma} P(\Gamma)\, {\rm H}_\Gamma
$. This effectively amounts to average the expectation values of
the gravitational variables in each box. We construct such
averages in terms of  the most general combinations  of flat space
tensors $\delta_{ab},  \epsilon_{abc}, \dots $ which saturate the
tensor structure of the classical fields together with their
derivatives in each box. In this way we are imposing rotational
invariance on our final effective Hamiltonian.

Next we make some general comments regarding the above procedure:
(i) our calculation has been performed in a fixed reference frame
and  leaves undetermined an overall numerical dimensionless
coefficient in each of the calculated contributions. The results
can be viewed as an expansion in terms of the classical fields and
their derivatives, combined with an explicit dependence upon the
two scales $\ell_P$ and $L$. (ii) The corrections obtained in this
way have been usually interpreted as signaling a preferred
reference frame together with a violation of the standard active
(particle) Lorentz transformations. The advent of DSR opens up the
possibility to study whether or not such modified actions can be
embedded in a related framework, thus recovering a modified
relativity principle.

\section{OBSERVATIONAL BOUNDS FOR FERMIONS USING EXISTING DATA}

Here we have taken the point of view that the previously found
Hamiltonians  account for the corresponding dynamics in a
preferred reference frame, which we identify as the one where the
Cosmic Microwave Background looks isotropic. Our velocity $\mathbf
w$ with respect to that frame has already been determined to be
$w/c\approx 1.23 \times 10^{-3}$ by COBE. Thus, in  the earth
reference frame one expects the appearance of signals indicating
minute violations of space isotropy encoded in $\mathbf
w$-dependent terms appearing in the transformed Hamiltonian or
Lagrangian. On the other hand, many high precision experimental
test of rotational symmetry, using atomic and nuclear systems,
have been already reported in the literature. Amazingly such
precision is already enough to set very stringent bounds on some
of the parameters arising from the quantum gravity corrections. We
have considered the case of non-relativistic Dirac particles,
obtaining corrections which involve the coupling of the spin to
the CMB velocity together with a quadrupolar anisotropy of the
inertial mass.\cite{boundsurru} The calculation was made with the
choices $\Upsilon=0$ and ${ L}=1/M$, where $M$ is the rest mass of
the fermion. Keeping only terms linear in $\ell_P$, the equation
of motion  arising from the two-component Hamiltonian obtained by
Alfaro, Morales-T\'ecotl and Urrutia \cite{URRUF} can be readily
extended to the Dirac case, yielding a Lagrangian which describes
the time evolution as seen in the CMB frame. In order to obtain
the dynamics  in the laboratory frame we implement an observer
Lorentz transformation by introducing explicitly the CMB frame's
four velocity $W^{\mu }=\gamma (1,\,{\mathbf{w}}/c)$. The result
is\cite{boundsurru}
\begin{eqnarray}
L_{D}&=&\frac{1}{2}i\bar{\Psi}\gamma ^{\mu }\partial _{\mu }\Psi -\frac{1}{2}M%
\bar{\Psi}\,\Psi +\frac{1}{2}i(\Theta_1 M\ell
_{P})\bar{\Psi}\gamma _{\mu }\left(
g^{\mu \nu }-W^{\mu }W^{\nu }\right) \partial _{\nu }\Psi \nonumber \\
&&+\frac{1}{4}(\Theta_2M\ell_{P})\bar{\Psi}\epsilon _{\mu \nu
\alpha \beta }W ^{\mu }\gamma ^{\nu }\gamma ^{\alpha }\partial
^{\beta }\Psi - \frac{1}{4}(\Theta_4 M\ell_P)M W_\mu
\bar{\Psi}\gamma_5\gamma^\mu\Psi +h.c.\, . \nonumber \\
\label{otra}
\end{eqnarray}
Here $\Theta_i,  \, i=1,2,4 $ are undetermined constants
parameterizing the dynamical corrections, expected  to be of order
one, and which can be identified in terms of the corresponding
Lorentz invariance violating tensors of the standard model
extension. From the work of Kosteleck\'y and Lane\cite{Lane} we
obtain the non-relativistic limit of the Hamiltonian corresponding
to (\ref{otra}), up to first order in $\ell_P$ and up to order
${({\bf w})/c}^{2}$. The relevant corrections are \be \delta
H_{S}= \left(\Theta_2 +\frac{1}{2}\Theta_4\right) M\ell _{P}
(2Mc^2) \left[ 1 + O\left( \frac{p^{2}}{2M^2c^2}\right) \right]
\mathbf{s}\cdot \frac{\mathbf{w}}{c}\,, \ee
\begin{equation}
\delta H_{Q}=-\Theta_1 M\ell _{P}\frac{5}{3}\left\langle \frac{p^{2}}{2M}%
\right\rangle \left( \frac{Q}{R^{2}}\right) \left(
\frac{w}{c}\right) ^{2}P_{2}(\cos \theta )\,. \label{QMM}
\end{equation}
The first represents a coupling of the nucleon spin $\mathbf{s}$
to the velocity of the CMB frame, while the second represents an
anisotropy of the inertial mass. Both contributions have been
bounded in Hughes-Drever like experiments. Here $Q$ is the
electric quadrupole moment and $R$ is the radius of the nucleus
under consideration, while $\theta$ is the angle between the
quantization axis and $\mathbf{w}$. Using $<p^{2}/2M>\sim 40$ MeV
for the energy of a nucleon in the last shell of a typical heavy
nucleus, together with the experimental bounds of Chupp et. al.
and Bear et. al.\cite{CHUBEAR} we find\cite{boundsurru}
\begin{equation}
\mid \Theta_2+\frac{1}{2}\Theta_4\mid <2\times 10^{-9},\qquad \mid
\Theta_1\mid <3\times 10^{-5}. \label{Result2}
\end{equation}
 The above  bounds  on quantities that were expected to be of
order unity already call into question the scenarios inspired on
the various approaches to quantum gravity, suggesting the
existence of Lorentz violating Lagrangian corrections which are
linear in Planck's length. In relation to this point it is
interesting to notice that a very reasonable agreement with the
current AGASA  ultra high energy cosmic ray (UHECR) spectrum
beyond the GZK cutoff has been recently obtained by using
dispersion relations of order higher than linear in $\ell_P$,
together with  considerations of additional stringent bounds
arising from  the first estimations of the impact of nearby BL Lac
objects and UHECR data upon LQG parameters.\cite{AP2002}

\section*{Acknowledgments}

The author acknowledges V.A. Kosteleck\'y for the invitation to
participate in the Focus Session on CPT and Lorentz symmetry of
QTS3. Partial support from  the projects CONACYT-40745-F and
DGAPA-IN11700 is also acknowledged.

\end{document}